# Classical to Quantum Software Migration Journey Begins: A Conceptual Readiness Model

Muhammad Azeem Akbar[1], Saima Rafi[2], Arif Ali Khan[3]

[1]Department of Software Engineering, LUT University, Lappeenranta, Finland
[2]University of Murcia, Department of Informatics and Systems, Murcia, Spain
[3]M3S Empirical Software Engineering Research Unit, University of Oulu, 90570 Oulu, Finland
`azeem.akbar@lut.fi, saeem112@gmail.com, arif.khan@oulu.fi`

**Abstract.** With recent advances in the development of more powerful quantum computers, the research area of quantum software engineering is emerging. Quantum software plays a critical role in exploiting the full potential of quantum computing systems. As a result, it has been drawing increasing attention recently to provide concepts, principles, and guidelines to address the ongoing challenges of quantum software development. The importance of the topic motivated us to voice out a call for action to develop a readiness model that will help an organization assess its capability of migration from classic software engineering to quantum software engineering. The proposed model will be based on the existing multivocal literature, industrial empirical study, understanding of the process areas, challenging factors and enablers that could impact the quantum software engineering process. We believe that the proposed model will provide a roadmap for software development organizations to measure their readiness concerning to transformation from classic to quantum software engineering by suggesting best practices and highlighting important process areas, challenges, and enablers.

## 1 Introduction

Quantum computing promises to solve many problems more precisely than possible with classical computers, e.g., simulating complex physical systems or applying machine learning techniques[1, 2]. Presently, that quantum computing has become widespread in developing more powerful quantum computers, and their need in terms of quantum software and applications, development process and frameworks, quantum software architectures and styles are becoming increasingly important [3, 4]. Quantum computing is a technological revolution that demands a new software engineering paradigm to develop and conceive quantum software systems. Quantum software engineering calls for novel techniques, tools, processes, and methods that explicitly focus on developing software systems based on quantum mechanics[5]. Though, the development of such quantum applications is complex and requires experts with knowledge from various fields, e.g., physics, mathematics, and computer science [6, 7].

Quantum software engineering is an emerging research area investigating concepts, principles, and guidelines to develop, maintain, and evolve quantum applications [8, 9]. Therefore, it is important to enhance the quality and reusability of the resulting quantum applications by systematically applying software engineering principles during all development phases, from the initial requirement analysis to the software implementation [10]. In classical software engineering, software development processes often document the different development phases a software artefact or application goes through [11, 12]. Furthermore, such software development process also summarizes best practices and methods that can be applied in the various phases and corresponding tools [9, 13]. Hence, they can be used for educating new developers by providing an overview of the development process or serving as a basis for cooperating with experts from different fields [14]. Today's quantum applications are often hybrid, consisting of quantum and classical programs [15]. Thus, the development process for quantum applications involves developing and operating both kinds of programs. However, existing lifecycles from classical software engineering [16] and quantum software development process [13, 17] only target one of these kinds and do not address the resulting integration challenges.

Furthermore, the execution of the quantum and classical programs must be orchestrated, and data has to be passed between them[18]. The workflow process is a means for these orchestrations to provide benefits, such as scalability, reliability, and robustness [19]. Thus, to transform from classic to quantum software development process, we need to analyze the software development community's tools, standards, and guidelines. Stefano et al. [20] highlighted that *"the challenge of quantum software engineering is to rework and extend the whole of classical software engineering into the quantum domain so that programmers can manipulate quantum programs with the same ease and confidence that they manipulate today's classical programs."* Ahmad et al. [21] presented the architectural view of quantum software engineering architecture (Fig 1). The presented architecture view helps to reflect in designing and envisioning an overall system, to avoid errors and future bugs in quantum system. Hence, the role of architecture is empowered in quantum software applications to abstract complexities of source code modules and their interactions as architectural component and connectors [22].

**Motivation scenario**

Despite the significance of quantum software engineering, no standards and models are available to handle quantum software development processes. For example, if an organization want to transform from classic to quantum software development, they need guidelines and strategies to put the process on the right path. Thus, it is required to estimate all aspects of a software development process like time, cost, integration aspect, scope, quality, human resources, risk, communication, stakeholders, and procurements. The transformation from classic to the quantum system is a challenging exercise due to issues such as:

***Little research*** has been conducted on the development of models and strategies. The problems faced by organizations during the implementation of quantum software development activities are quite different from the traditional or classical paradigm. Therefore, existing literature doesn't examine the transformation from classic to quantum software engineering in sufficient detail as there is little research that highlights the important process areas and challenges to address for the adoption of quantum software development. Therefore, ***lack of proper guidelines*** that help practitioners to implement quantum technology for software development. Presently, there are no assessment tools and frameworks for determining an organization's readiness concerning transforming from a classic to a quantum software development process. No such practices are available that assist practitioners in improving quantum software engineering in their organization.

Moreover, there is a ***lack of a roadmap*** to help organizations choose the appropriate patterns, particularly for their problems. No study ***addresses the project management changes*** caused due to the migration from classic to quantum software engineering. Thus, it is demanded to deeply study the important process areas, challenges, enablers, and guidelines that could influence the adoption of quantum software development. Furthermore, discussing the different software artefacts usually constituting a quantum application and presenting their corresponding process areas is required. It is ***critical to identify the plug points*** between the classic and quantum software modules to enable their integration into overall application, for execution of hybrid quantum applications. To address all the highlighted concerns, there is need of practically robust roadmap and guidelines to assist the practitioners to make the migration from classic to quantum software development successful. Hence, the readiness model is one of the key instruments to assists software development organizations to assess the capability of an organizations concerning to transform from classic software engineering to quantum software engineering.

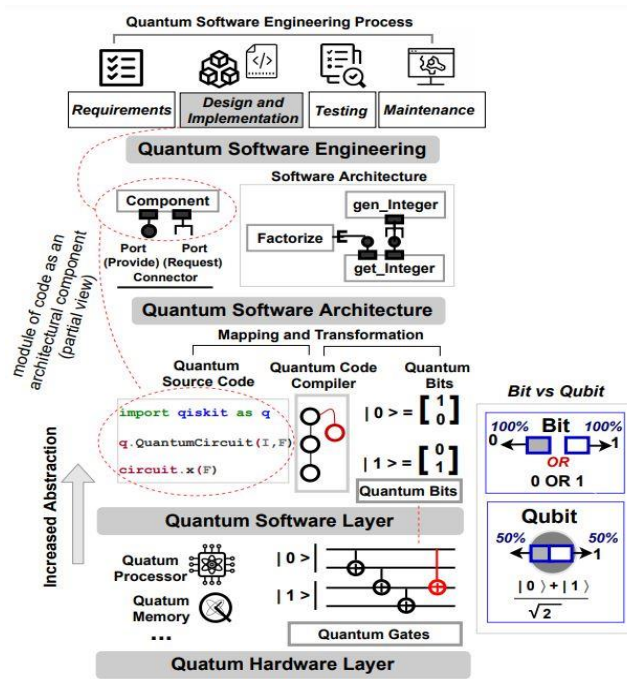

**Fig. 1**. Architecture of quantum software engineering [21]

**Readiness Models and Standards**
A readiness model is a technique to assess an organization or team based on the specified criteria to represent their level of readiness. Readiness models are intended to help organizations appraise their process readiness and develop it for improvement. They serve as points of reference for different stages of readiness in an area. Software engineering readiness models intend to help organizations move from ad-hoc processes to mature and disciplined software processes[23].

In software engineering research, a readiness model has been utilized in several studies. It was used by Niazi et al. [24] to assess organizational readiness in terms of software process improvement. Their readiness model has several levels: aware, defined, and optimizing. Critical factors and barriers support each level. The researchers validated their readiness model by performing case studies in three software organizations. Similarly, Ali and Khan[25] presented a model to measure the readiness of a software organization to form outsourcing relationships. They utilized critical partnership factors to develop a readiness model and examined their practical implementation. Their readiness model has several levels: contract, success, readiness, conversion and maturity. Similarly, Khan et al.[26] proposed a software outsourcing vendor readiness model (SOVRM). The readiness levels of the SOVRM consist of critical barriers and critical success factors. Similarly, a recent study conducted by Sufi et al.[27] proposed security requirements engineering readiness (SRERM). The levels of SRERM are based on security requirements categories. All the above-discussed readiness models followed the capability maturity model Integration (CMMI) staged representation structure and considered the critical barriers and success factors as the key process areas (KPA's). The software engineering institute developed CMMI almost twenty years ago[28].

CMMI helps organizations to streamline process improvement. It clearly shows what organizations should do to mature software processes. CMMI model is integrated into five maturity levels, i.e., (initial, managed, defined, quantitatively managed, and optimizing). CMMI had proved itself for decades yet has had no meaningful impact in providing detailed information about broader technology space such as quantum computing in implementing strategies and key practices.

ISO/IEC 15504 Information Technology: SPICE is an international framework for accessing software development [29]. It provides a detailed description of standard documents for the software development process and related management functions within an organization. It includes two dimensions, i.e., capability dimension and process dimension. It also introduces assessment indicators that help an organization with brief guidelines to assess the quality of their management process. To see in terms of improving quantum computing process areas, SPICE does integrate existing process improvement methodologies. Still, it does not provide an explicit process improvement path regarding quantum software.

International Standards Organization (ISO) 9000/9001: ISO 9000 is a series of standards in quality management that helps organizations maintain their customer and other stakeholder needs related to a product or service [30]. It helps organizations to document the elements needed for quality software systems effectively. ISO 9001 consists of generic standards that are not specific to the only software industry and can be applied to different types of organizations. These standard guidelines focus on the industry's manufacturing and services aspects, including quality standards. However, it still lags behind process improvement aspects of software systems while using quantum technology.

Several readiness and maturity models have been proposed by researchers and practitioners in the traditional software development domain, providing a framework to assess an organization's current effectiveness and supporting figuring out what capabilities they need to acquire next to improve their performance. Indeed, this apparent popularity of these models out on the field has partly motivated us to propose a readiness model in the context of transformation from classic to quantum software development. In an area where we struggle with a gap between research and practice, we argue that looking at frameworks, models, and other tools actively used out on the field is a good starting point for further steps. Thus far, guidelines have been used to make quantum software engineering more tangible, but further steps are still needed, and a robust readiness model could be one such step.

## 2 Call for Action

We propose developing a readiness model to provide a roadmap for migrating from classical to quantum software development. Such a readiness model would help the field move from ad hoc implementation of quantum software development to a more mature process. Furthermore, we argue that this model should not be an effort for a single researcher or research group but a multidisciplinary project that builds on a combination of theoretical models and empirical results. The research work is classified in four steps to developing the proposed readiness model.

**Step 1:** This step will give a broad overview of the available literature and identify the key process areas and challenging factors that can influence the transformation from classic to quantum software development process. To meet this objective, we plan to conduct a multivocal literature review (MLR) which is a viable approach to extracting data from the grey and white literature.

As the topic under investigation is not maturely studied in mainstream research, thus the grey literature could give critical insights about it. The key finding of this step revolves around the following questions.

*[What process areas of transformation from classic software development to quantum software development are reported in the existing literature?]*
*[What are the key challenging factors of transforming the classic software development process to quantum, reported in the literature?]*
*[What enablers are essential for transforming the existing classic to quantum software development process, reported in the literature?*

**Step 2:**
This step leads to empirically validating the literature findings (Steps 1) with industry practitioners by conducting the questionnaire survey, case study, and interviews. This step aims to confirm significant process areas and challenges identified in step 1 and to enlist additional influencing areas towards transforming the traditional software development process into quantum. In this step, we will find the answers to the following questions:

*[What process areas are critical to consider while transforming from classic to quantum software development process?]*
*[What are the key challenges faced by industrial practitioners while transforming the existing classic software development process to quantum software development?]*
*[What enablers are essential for transforming the existing classic to quantum software development process? discussed in real-world practice?]*

**Step 3:** This step will investigate best practices against each identified challenging factor and enabler (in Steps 1 and 2). To achieve this step, we will conduct MLR to investigate the state-of-the-art best practices reported in grey and formal literature. Furthermore, we will empirically explore the best practices against each challenging factor and enabler by conducting a questionnaire, case study, and interviews. This step will answer the following questions:

*[What best practices address the challenging factors (Step 1), reported in the literature and real-world industry?]*
*[What are the best practices to achieve the enablers identified in Step 2, reported in the literature and real-world industry?]*

**Step 4:** Finally, a readiness model will be developed to assist the software development organizations in assessing, adapting and improving their process toward the migration from classic to quantum software development paradigm. To develop the readiness model, we will consider the findings of steps 1, step 2, and steps 3.

The readiness model will consist of three components, i.e., the assessment component, factors component (process areas, challenges, enablers), and guidelines component. The identified best practices will be mapped against each enabler and challenging factor to achieve that certain level. If an organization wants to move to the next level, they need to address each enabler and challenging factor by implementing its respective best practices.
The developed readiness model will help the organizations assess their ability with respect to the transformation from classic to quantum software development and provide a roadmap to improve their capability concerning the adoption of quantum software development.
To check the practical robustness of the model, we will conduct case studies in software development organizations and update them according to their suggestions. The final model will be available for software development organizations to adopt and improve their adaptability and executability concerning to quantum software development process.

*[How to develop and evaluate the effectiveness of the proposed model?]*
*[What would be the readiness levels of the proposed model?]*
*[How to check the robustness of the proposed model in the real-world industry?]*

## 3　Architecture of proposed model

The basic architecture of the proposed quantum software engineering readiness model (QSERM) will be designed based on Process areas and their associated challenges and key enablers identified from literature and industry practices. To align identified components in the structured model, we will use the concept of existing software engineering standards such as CMMI, IMM and SPICE. Fig. 2 shows the relationship between key components of the proposed model. It depicts the proposed model's complete component, highlighting how the results of existing models, literature and industry findings will be used to design the key components of the proposed QSERM.

The four components of QSERM are:
- Readiness level component
- Process areas
- Challenge
- Key enablers

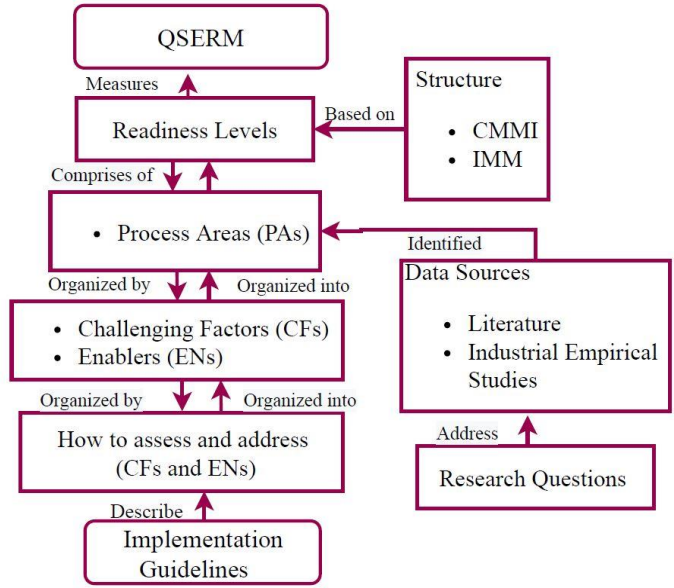

**Fig. 2.** Structure of the proposed model

### 3.1 Readiness level component

The proposed model consists of readiness levels based on the standard model for software engineering i.e., CMMI. Several adjustments are required in the structure of CMMI to make it applicable for quantum software applications. The structure of each readiness level is given in Fig. 3, and brief explanation is given below:

**Process Areas (PAs):** Process areas are the building blocks that indicate the areas an organization should focus on to improve software processes. These areas consist of a cluster of related practices that when implemented collectively, satisfy the goals related to that area. Therefore, we will identify the process areas related to quantum software engineering to improve the software development process.

**Challenging Factors (CFs):** The architecture of the proposed model consists of various process areas. The identified challenging factors will be mapped to all maturity levels and process areas associated with each level. This formulation has been used previously by many researchers. Therefore, we can justify the use of challenging factors in our study.

**Enablers (ENs):** The Key enablers will be identified to support the proposed model to accomplish the goals associated with all five maturity levels of QSERM. To justify the use of key enablers, it provides the best support to perform essential tasks. We will perform an SLR study to identify the key enablers from software engineering experts working with quantum development.

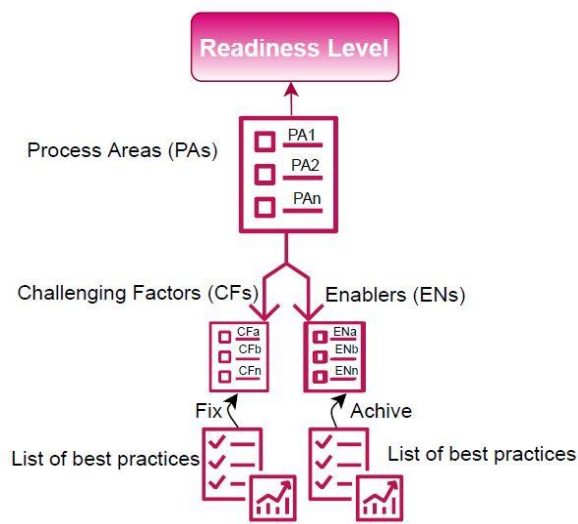

**Fig. 3.** Structure of each readiness level

The proposed QSERM will be based on five readiness levels (Fig. 4). Each readiness level encompasses specific process areas. The process areas highlight the important zones that need to be addressed by an organization. Furthermore, important, challenging factors and enablers will be aligned with each process area. To achieve a higher level, an organization must address all the process areas of a readiness level. And to address all the process areas, organizations must address all the challenging factors and enablers. The best practices will be mapped against each challenge and enabler, which will assist the organizations in addressing them effectively. For example, if organization-A wants to move to level 2, they need to address all the process areas of level 1. To achieve this, they need to address all the challenging factors and enablers of level-1 by implementing their associated best practices.

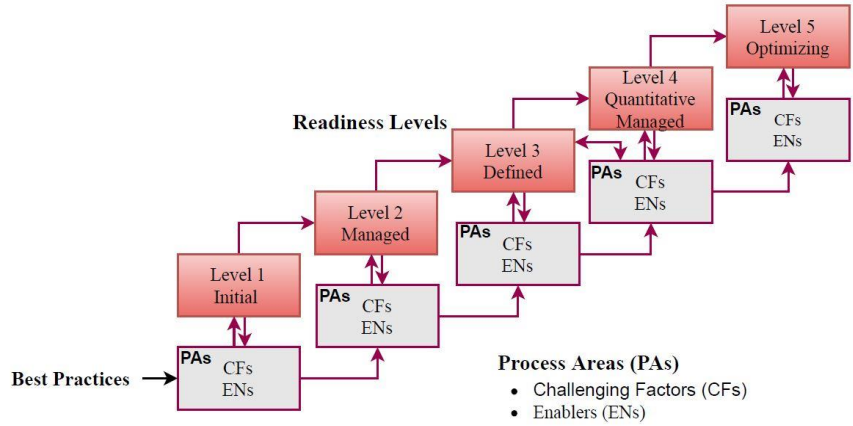

**Fig. 4.** Example of proposed readiness model

### 3.2 Assessment Dimension

To evaluate the model, we will use the Motorola assessment tool [31]. Many researchers in software engineering field have used this tool to evaluate their proposed readiness model. Therefore, we have selected the same tool for the evaluation of QSERM. This tool will assist the organization in identifying the areas that need further improvement. The three dimensions of the Motorola assessment tool are:

**Approach:** Emphasize the top management's commitment to implementing the specific practice.
**Deployment:** Focus on the consistent and uniform practice implementation across quantum project areas.
**Results:** Assess the breadth and consistency of the results of deployed practice across different project areas.

## 4 Expected outcomes

Since in the early stage, the study will highlight only a few contributions. One of the contributions is identifying process areas, challenges, enablers, and associated practices that will help quantum software development. The process areas consist of a cluster of related practices that, when implemented collectively, satisfy the goals related to that area. The second contribution is to develop a quantum software engineering readiness model. This model will assist organisations in assessing readiness and suggest guidelines for successfully adopting the quantum software engineering paradigm. And the third contribution is to help organizations in "identifying", "analyzing" and "mitigating" the challenges faced during the migration from classic to quantum software engineering. The novelty of this research work is the development of a readiness model that will state activities, guidelines or roadmap that can be assist in migrating from classic to quantum software development.